\documentclass[prb,superscriptaddress,twocolumn,reprint,showpacs]{revtex4}
\usepackage{amsmath, amsthm, amssymb}
\usepackage{graphicx}
\usepackage{dcolumn} 
\usepackage{bm} 
\makeatletter 
\gdef\@ptsize{2}
\let\@currsize\normalsize 
\makeatother 
\usepackage{setspace} 

\newcommand {\BIO}{Ba$_{2}$IrO$_{4}$}
\newcommand {\SIO}{Sr$_{2}$IrO$_{4}$}
\newcommand {\SRO}{Sr$_{2}$RhO$_{4}$}
\newcommand {\SCO}{Sr$_{2}$CoO$_{4}$}
\newcommand {\LCO}{La$_{2}$CuO$_{4}$}

\newcommand {\LSBIOd}{La$_{x}$Sr$_{2-x-y}$Ba$_{y}$IrO$_{4-\delta}$}

\newcommand {\LSIOd}{La$_{x}$Sr$_{2-x}$IrO$_{4-\delta}$}
\newcommand {\GSO}{GdScO$_{3}$}
\newcommand {\DSO}{DyScO$_{3}$}
\newcommand {\STO}{SrTiO$_{3}$}
\newcommand {\ttwog}{$t_{2\mathrm{g}}$}
\newcommand {\Jhalf}{$J_{\mathrm{eff}}=1/2$}
\newcommand {\Shalf}{$S=1/2$}
\newcommand {\Jthreehalf}{$J_{\mathrm{eff}}=3/2$}

\usepackage[usenames,dvipsnames]{color}

\begin{document}
\title{Effective Carrier Doping and Metallization in {\LSBIOd} Thin Films}

\author{M. Ito}
\affiliation{Institute for Materials Research, Tohoku University, Sendai 980-8577, Japan}
\affiliation{RIKEN Center for Emergent Matter Science (CEMS), Wako 351-0198, Japan}
\author{M. Uchida}
\email[Author to whom correspondence should be addressed: ]{uchida@ap.t.u-tokyo.ac.jp}
\affiliation{Quantum-Phase Electronics Center and Department of Applied Physics, University of Tokyo, Bunkyo-ku, Tokyo 113-8656, Japan}
\author{Y. Kozuka}
\affiliation{Quantum-Phase Electronics Center and Department of Applied Physics, University of Tokyo, Bunkyo-ku, Tokyo 113-8656, Japan}
\author{K. S. Takahashi}
\affiliation{RIKEN Center for Emergent Matter Science (CEMS), Wako 351-0198, Japan}
\author{M. Kawasaki}
\affiliation{RIKEN Center for Emergent Matter Science (CEMS), Wako 351-0198, Japan}
\affiliation{Quantum-Phase Electronics Center and Department of Applied Physics, University of Tokyo, Bunkyo-ku, Tokyo 113-8656, Japan}

\date{\today}


\begin{abstract}

We fabricate {\LSBIOd} thin films by pulsed laser deposition, in an effort to realize the effective carrier doping and metallization in the {\SIO} system. We design ideal in-plane Ir-O-Ir frame structure by utilizing tensile substrate strain and Ba substitution, as well as control La doping and oxygen deficiency. This enables us to elucidate relation between the charge transport and the carrier density through systematic changes from original $p$-type spin-orbit Mott insulator to highly doped $n$-type metal.

\end{abstract}
\pacs{73.50.-h, 71.30.+h, 71.70.Ej}
\maketitle

\section{Introduction}

Transition metal oxides have attracted growing interest due to their emergent phenomena induced by carrier doping, such as high-temperature superconductivity and colossal magnetoresistance. These phenomena originate from mutual and collective interactions among charge, spin, and orbital degrees of freedom of the $d$ electrons, which become dominant near the Mott metal-insulator transition. \cite{MIT} Recently, 5$d$ transition metal oxides have been intensively studied in the light of strong spin-orbit coupling. In the 5$d$ transition metal oxides, the energy scale of the spin-orbit coupling is typically comparable to that of the Coulomb interaction. This gives rise to an unconventional electronic configuration, which has been confirmed in {\SIO}, a 5$d^5$ model system. \cite{ito5, ito6} For example, if the spin-orbit coupling were negligibly weak in {\SIO}, a metallic ground state would have been expected to appear due to the itinerant character of wide 5$d$ bands. Actually, {\SRO} and {\SCO}, 4$d^5$ and 3$d^5$ counterpart systems, are a paramagnetic and ferromagnetic metals, respectively. \cite{ito2, ito3} However, {\SIO} is known to be an antiferromagnetic insulator. \cite{ito4} Recent studies have revealed that the strong spin-orbit coupling results in a so-called spin-orbit Mott insulator state with effective total angular momentum {\Jhalf}, \cite{ito5, ito6} where the $5d^{5}$ (Ir$^{4+}$) {\ttwog} orbitals are split into a fully filled {\Jthreehalf} manifold and a half-filled {\Jhalf} state. Due to the Coulomb interaction, the half-filled {\Jhalf} band is gapped and further split into an occupied lower band and an unoccupied upper band.

In the sense that the half-filled pseudospin {\Jhalf} state is analogous to the {\Shalf} one, {\SIO} is similar in electric configurations to {\LCO}, a parent compound of the high-temperature superconductors. \cite{ito8} The superconductivity emerges in layered cuprates by carrier doping into the two-dimensional {\Shalf} antiferromagnetic ordering. Since {\SIO}, having the layered perovskite structure, is also isostructural to {\LCO}, it is intriguing to investigate carrier doping effects on the {\Jhalf} case. \cite{siosc} Recently, it has been theoretically predicted that superconductivity may be realized by doping electrons or holes into the layered iridates. \cite{ito7, ito9, hae} Numbers of groups have actually investigated the carrier doping aiming at the superconductivity in {\SIO} or a sister compound {\BIO}, \cite{BIO1, BIO2, BIO3, BIO4} for example by substituting Sr or Ba with La or K atoms \cite{ito12, ito13, ito14, BIOdoping} or introducing O deficiency. \cite{ito10, ito11} Carrier injection by electric field effect has been also tried using single-crystalline thin films. \cite{ito16, ito17} However, no superconductivity has been reported so far, and even metallization is rather difficult and poorly organized in this system.  

In this paper, we focus on designing ideal in-plane Ir-O-Ir frame structure with using thin films, in order to effectively doping carriers into the {\SIO} system. For this purpose, we utilize tensile substrate strain and Ba substitution, as well as control the La doping and O deficiency. Changes from $p$-type spin-orbit Mott insulator to highly doped $n$-type metal are then systematically observed in the transport measurement.

\section{Experiment}

{\LSBIOd} ($x=0$, 0.1, 0.2, and $y=0$, 0.6, 0.9, 1.2) thin films were epitaxially grown by pulsed laser deposition, on single-crystalline (LaAlO$_3$)$_{0.3}$-(Sr$_2$AlTaO$_6$)$_{0.7}$ (LSAT), {\STO}, {\DSO}, and {\GSO} substrates. Polycrystalline {\LSBIOd} targets were prepared by a spark plasma sintering method. The sintering was performed at 950 $^{\circ}$C in 0.05 MPa Ar atmosphere with applying mechanical pressure of 80 MPa. For deposition, a KrF eximer laser ($\lambda=248$ nm) was used to ablate the targets. Laser fluence and repetition rate were set to 0.5 J/cm$^{2}$ and 10 Hz, respectively. Films were deposited at a substrate temperature of 800 $^{\circ}$C in 10 mTorr O$_{2}$. After the growth, samples were cooled down in 10 mTorr O$_{2}$ or vacuum atmosphere, at a rate of 30 $^{\circ}$C/minute. Film thicknesses were about 30 nm for all the samples. Crystal structure was characterized using x-ray diffractometer (SmartLab, Rigaku Co.) with a source of CuK$_{\alpha 1}$. Longitudinal and Hall resistivities were measured using a van der Pauw method, in a $^{4}$He cryostat equipped with a 9 T superconducting magnet (PPMS, Quantum Design Co.).

\section{Results and discussion}

\subsection{Strategy}

\begin{figure}
\begin{center}
\includegraphics*[width=8.4cm]{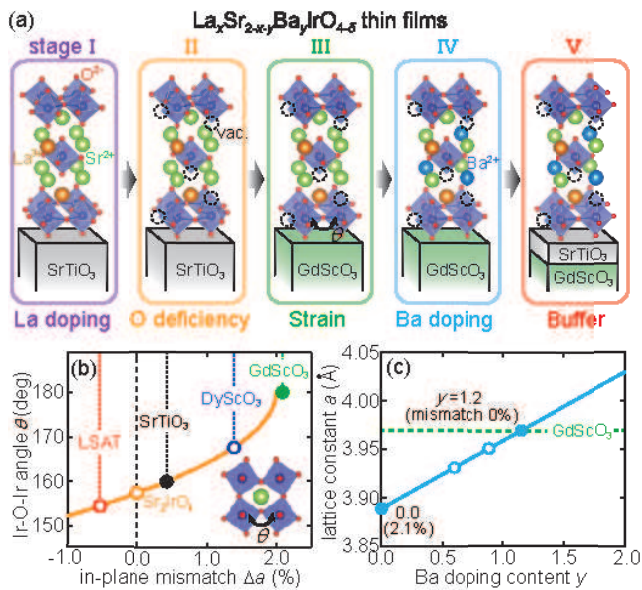}
\caption{(Color online)
(a) Growth strategy of {\LSBIOd} thin films towards effective carrier doping into the {\SIO} system. La doping (stage I), introduction of oxygen deficiency (II), strain control (III), Ba doping (IV), and insertion of buffer layer (V) are sequentially performed at each stage. (b) Relation between the in-plane lattice mismatch $\Delta a$ and the Ir-O-Ir bond angle $\theta$ is evaluated for various substrates under the assumption that the Ir-O bond length is constant, for realizing the straight bond angle at the stage III. (c) In-plane lattice constant $a$ is calculated against the Ba doping content $y$ with using Vegard's law, for matching the lattice constant to {\GSO} one at the stage IV.
}
\label{fig1}
\end{center}
\end{figure}

In order to effectively metalize the {\SIO} system, thin films were systematically fabricated following a roadmap consisting of stages I to V, as summarized in Fig. 1(a). At the stage I, {\LSIOd} ($x = 0$, 0.1, 0.2, $\delta < 0$) films were grown on the (001) {\STO} substrates, followed by cooling down in the original growth atmosphere of 10 mTorr O$_{2}$. For more effective electron doping, after the stage II, oxygen deficiency is intentionally introduced ($\delta > 0$) by cooling down in the vacuum after growth.

The in-plane Ir-O-Ir bond angle is also worth considering for further metallization. In bulk {\SIO}, IrO$_{6}$ octahedra are rotated 11$^{\circ}$ with respect to the $a$-axis. \cite{ito18, ito19} Such bent Ir-O-Ir bond may suppress the conduction on the IrO$_{2}$ plane. At the stage III, {\LSIOd} ($x = 0$, 0.1, 0.2) thin films were grown on various substrates, in an effort to realize the straight Ir-O-Ir bond by epitaxial strain. Figure 1(b) shows relation between the in-plane lattice mismatch $\Delta a$ and the Ir-O-Ir bond angle $\theta$. The in-plane mismatch $\Delta a$ is calculated for the {\SIO} bulk value with assuming that the Ir-O bond length is constant. Under this assumption, the {\GSO} substrate is favorable for designing the straight Ir-O-Ir bond.

After the stage IV, Ba is substituted for Sr site in order to match the in-plane lattice constant with {\GSO} one. At the stage IV, {\LSBIOd} thin films were grown on {\GSO}. Following the in-plane lattice constant calculated from the Vegard's rule, Ba doping contents $y = 0.6$, 0.9, and 1.2 are determined so that their mismatches to {\GSO} correspond to -1.0, -0.5, and 0\%, respectively, as shown in Fig. 1(c). In particular, $y=1.2$ is expected to realize the best lattice match with that of the {\GSO} substrate. At the stage V, finally, in an effort to avoid the valence mismatch at the interface and improve the film crystallinity, {\STO} buffer layer of about 5 nm was inserted between the {\LSBIOd} film and {\GSO} substrate.

\subsection{Structural characterization}

\begin{figure}
\begin{center}
\includegraphics*[width=8.4cm]{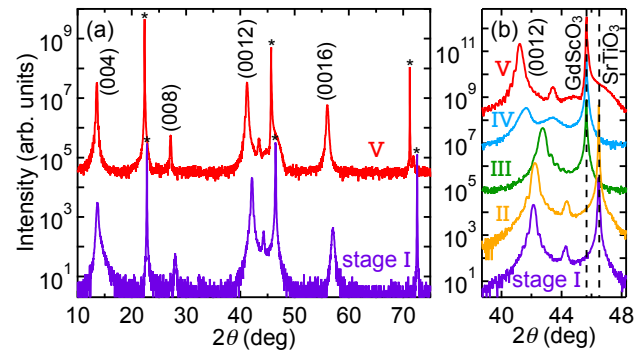}
\caption{(Color online)
(a) Typical x-ray diffraction $\theta$--2$\theta$ scans and (b) their magnified views around the (0012) peak for {\LSBIOd} films grown at the stages I to V. The film composition is $x=0.2$, $y=0.0$ for I, II, and III, and $x=0.2$, $y=1.2$ for IV and V. {\STO} and {\GSO} substrate peaks are marked with an asterisk.
}
\label{fig2}
\end{center}
\end{figure}

Figure 2(a) shows typical x-ray diffraction $\theta$--2$\theta$ scans of the thin films grown at the stages I and V. Sharp (004$l$) peaks ($l$: integer) are observed for both of the films, indicating that $c$-axis oriented films are epitaxially grown as illustrated in Fig. 1(a). In addition to these peaks, an unassigned peak is observed at around 44$^{\circ}$. This impurity phase is frequently observed in other layered perovskite films such as Sr$_2$RuO$_4$,\cite{sr2ruo4} and attributed to other Ruddlesden-Popper phase. The phase volume of $\sim$1\%, estimated from the peak intensity, is much lower than typical percolation thresholds, and thus this impurity phase should not dominate transport property.

Magnified views around (0012) peak in $\theta$--2$\theta$ scan are shown in Fig. 2(b) for the typical films grown at each stage. Film thicknesses obtained from the Laue fringes are about 30 nm. Comparing the stage II to I, the peak position is slightly shifted to higher angles, implying that the $c$-axis lattice constant decreases by introducing oxygen vacancy, which is opposite trend to other perovskite oxides. At the stage III, the peak position is shifted to much higher angle, indicating that the $c$-axis is elastically shortened due to the in-plane tensile strain from the {\GSO} substrate. After the stage IV, the $c$-axis gets longer by the substitution of Ba for Sr. Inserting the {\STO} buffer layer at the stage V dramatically improves the film crystallinity, compared to the broad film peak at the stage IV. 

\begin{figure}
\begin{center}
\includegraphics*[width=8.4cm]{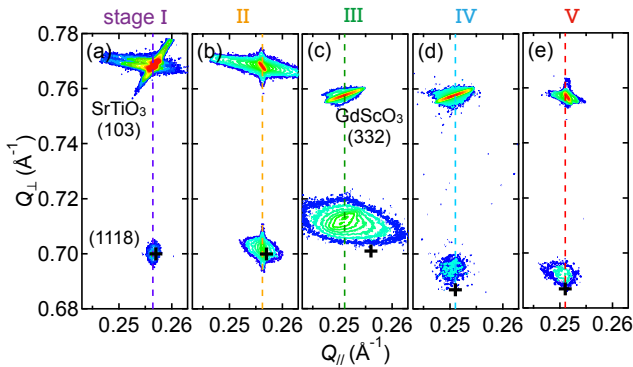}
\caption{(Color online)
(a)-(e) X-ray reciprocal space mappings around (1118) peak for the {\LSBIOd} films. The film composition at each stage is the same as those shown in Fig. 2. Vertical dashed lines represent the in-plane lattice match to the {\STO} and {\GSO} substrates. Crosses denote film peak positions calculated from the bulk lattice parameters. 
}
\label{fig3}
\end{center}
\end{figure}

Figure 3 summarizes the x-ray reciprocal space mappings for the same set of films shown in Fig. 2(b), which confirm relation of the in-plane lattice structure between the film and the substrate. The (1118) film peaks are clearly observed for all the films. The thin films grown at the stages I and II are strained to the {\STO} substrate with original lattice mismatch of 0.4\%, as shown in Figs. 3(a) and 3(b). The same composition film at the stage III are partially relaxed on {\GSO} with mismatch of 2.1\%, as seen in the broad peak in Fig. 3(c). On the other hand, Figures 3(d) and 3(e) show that the films are coherently grown on {\GSO} by optimizing the in-plane lattice constant with the Ba substitution of $y=1.2$ after the stage IV, even in the case of inserting the {\STO} buffer layer.

\subsection{Electrical characterization}

\begin{figure}
\begin{center}
\includegraphics*[width=7.0cm]{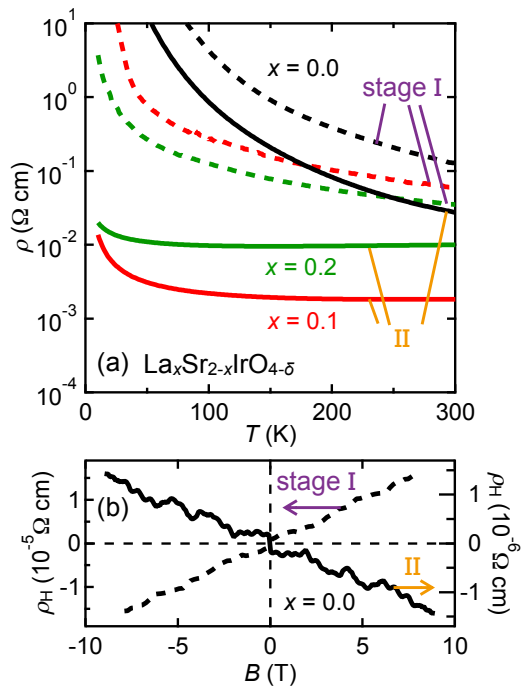}
\caption{(Color online)
(a) Temperature dependence of the in-plane resistivity of {\LSIOd} ($x=0.0$, 0.1, 0.2) films grown at the stages I (dashed lines) and II (solid lines). (b) Magnetic field dependence of the Hall resistivity for the films with $x=0.0$, measured at 300 K. The carrier type is obviously inverted from hole to electron, due to the oxygen deficiency introduced by vacuum cooling at the stage II. The carrier density is deduced to $3 \times 10^{20}$ $\mathrm{cm}^{-3}$ (I) and $4 \times 10^{21}$ $\mathrm{cm}^{-3}$ (II) in one carrier model.
}
\label{fig4}
\end{center}
\end{figure}

Following the above structural characterization, we evaluate transport properties of the films. Figure 4(a) shows temperature dependence of the in-plane resistivity for the films grown at the stages I and II. At the stage I, all the films including the La doped ones show insulating behavior. A carrier type of the non-doped film is hole, not electron, as shown in the Hall resistivity in Fig. 4(b). This indicates that excess oxygens of more than 10$^{20}$ cm$^{-3}$ are unintentionally incorporated and the La doping effect is fully compensated. In contrast, the in-plane resistivity of the La doped films grown at the stage II decreases with decrease of temperature and shows weak upturn at low temperature. Expectedly, the carrier type is reversed to electron even for the non-doped film, which is a firm evidence that the oxygen deficiency is effectively introduced during the vacuum cooling process. Reducing innate excess oxygen is of great significance for doping electrons into the {\SIO} system. 

\begin{figure}
\begin{center}
\includegraphics*[width=7.0cm]{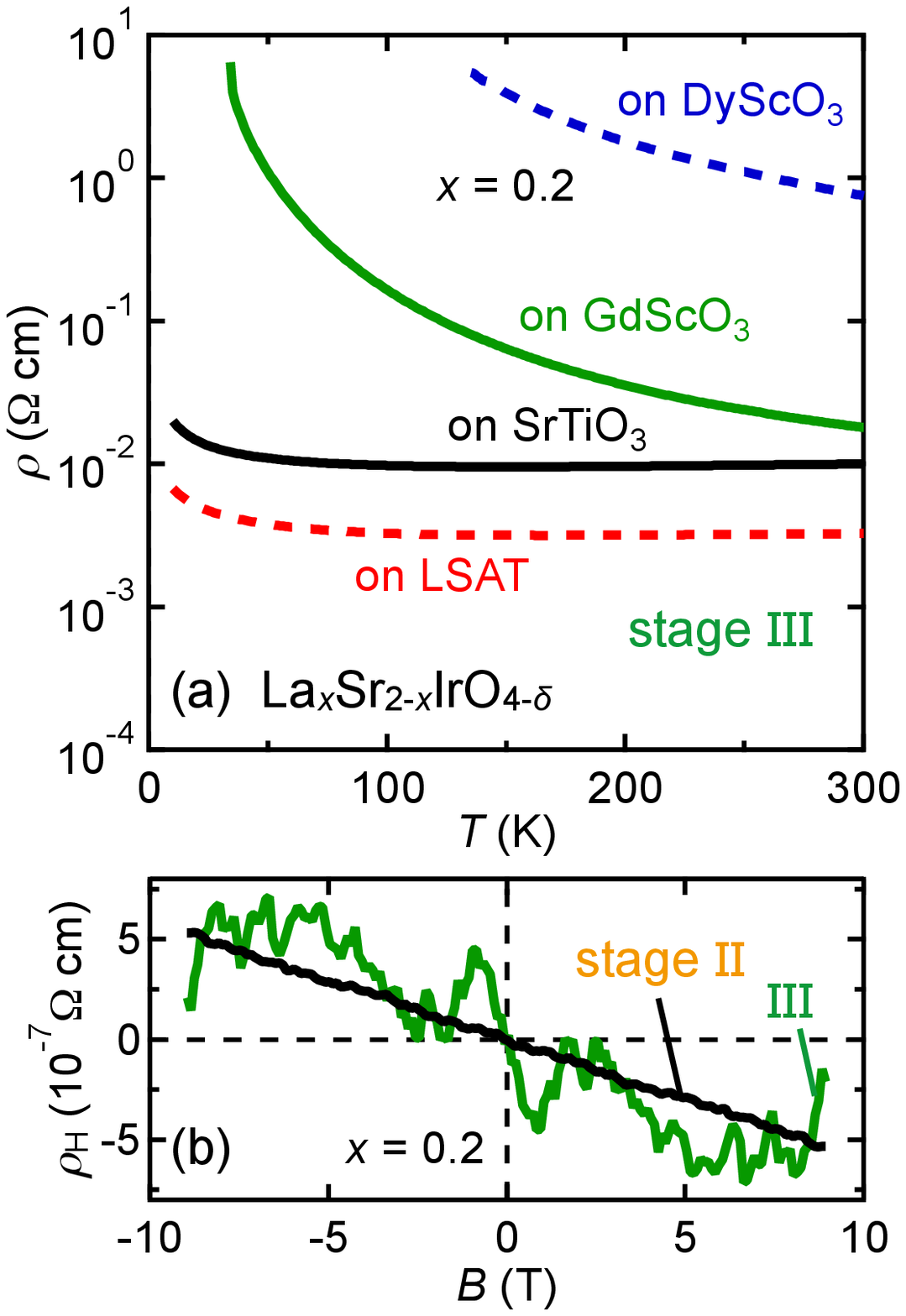}
\caption{(Color online)
(a) Temperature dependence of the in-plane resistivity of {\LSIOd} ($x=0.2$) films grown on various substrates at the stage III. The lattice mismatch to {\DSO} or {\GSO} substrate causes clear insulating behavior. (b) Magnetic field dependence of the Hall resistivity for the films grown on the {\STO} (black line) and {\GSO} (green line) substrates, measured at 300 K. The carrier density is deduced to $1 \times 10^{22} $ $\mathrm{cm}^{-3}$ (II and III) in one carrier model.
}
\label{fig5}
\end{center}
\end{figure}

Figure 5(a) shows temperature dependence of the in-plane resistivity for the films grown on various substrates at the stage III. As confirmed in Fig. 3(c), the {\LSIOd} ($x=0.2$) film grown on {\GSO} has a tensile strain, resulting in more straight Ir-O-Ir bond favoring the metallic conduction. However, the {\LSIOd} film on {\GSO} shows insulating behavior, in contrast to the films grown on {\STO} and LSAT. This is probably because numbers of dislocations originating from the 2.1\% lattice mismatch cause the strong electron scattering. As proof of this, both the films grown on {\STO} and {\GSO} are highly doped on the similar doping level, as confirmed in Fig. 5 (b).

\begin{figure}
\begin{center}
\includegraphics*[width=7.0cm]{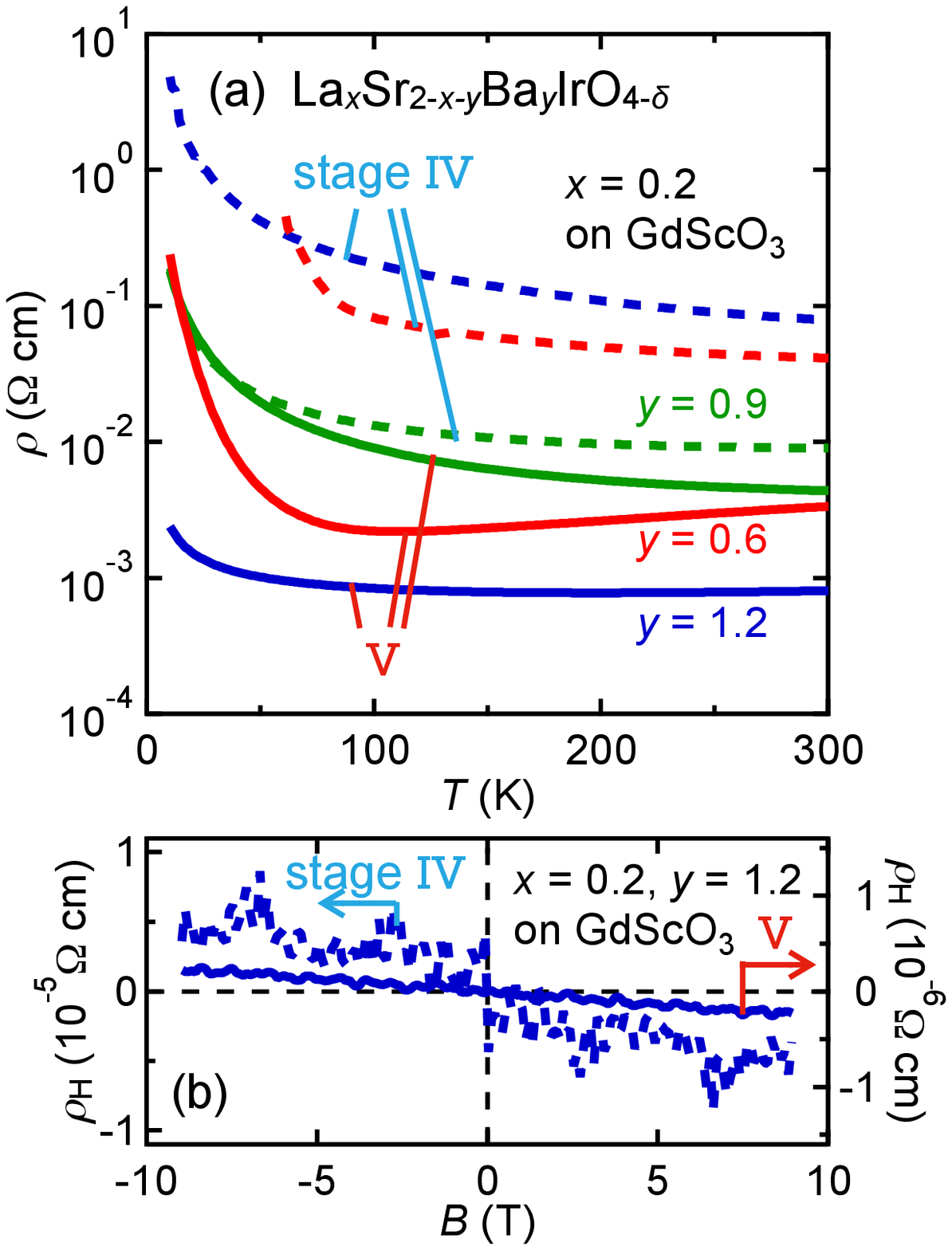}
\caption{(Color online)
(a) Temperature dependence of the in-plane resistivity for {\LSBIOd} ($x=0.2$, $y=0.6$, 0.9, 1.2) films grown at the stages IV (dashed lines) and V (solid lines). The resistivity is substantially reduced by inserting {\STO} buffer layer at the stage V. (b) Magnetic field dependence of the Hall resistivity for the films with $x=0.2$ and $y=1.2$, measured at 180 K. The carrier density is deduced to $2 \times 10^{21}$ $\mathrm{cm}^{-3}$ (IV) and $3 \times 10^{22}$ $\mathrm{cm}^{-3}$ (V) in one carrier model.
}
\label{fig6}
\end{center}
\end{figure}

Figure 6(a) shows the in-plane resistivity of the {\LSBIOd} films grown at the stages IV and V. Although the in-plane lattice constant is tuned to {\GSO} one by the Ba substitution, the films grown at the stage IV exhibits insulating behavior. On the other hand, inserting the {\STO} buffer layer at the stage V dramatically reduces the in-plane resistivity. The effect of {\STO} buffer layer may be interpreted to avoid valence mismatch at the heterointerface. Electric reconstruction usually occurs at the valence-mismatched perovskite interface, for example, between La$^{3+}$Al$^{3+}$O$_{3}$ and Sr$^{2+}$Ti$^{4+}$O$_{3}$.\cite{ito20} The formation of Sr$^{2+}_{2}$Ir$^{4+}$O$_{4}$ will be adversely affected from energetically unfavorable interface with Gd$^{3+}$Sc$^{3+}$O$_{3}$, whereas it will be avoidable by inserting Sr$^{2+}$Ti$^{4+}$O$_{3}$. Although another interface between Sr$^{2+}$Ti$^{4+}$O$_{3}$ and Gd$^{3+}$Sc$^{3+}$O$_{3}$ may suffer from the same, high-quality {\STO} buffer layer can be grown on the {\GSO} substrate,\cite{ito21} which is probably because {\STO} is thermodynamically much more stable. As shown in the Hall resistivity in Fig. 6(b), carrier type remains electron and high carrier density comparable to the stage II is realized.

\begin{figure}
\begin{center}
\includegraphics*[width=8.4cm]{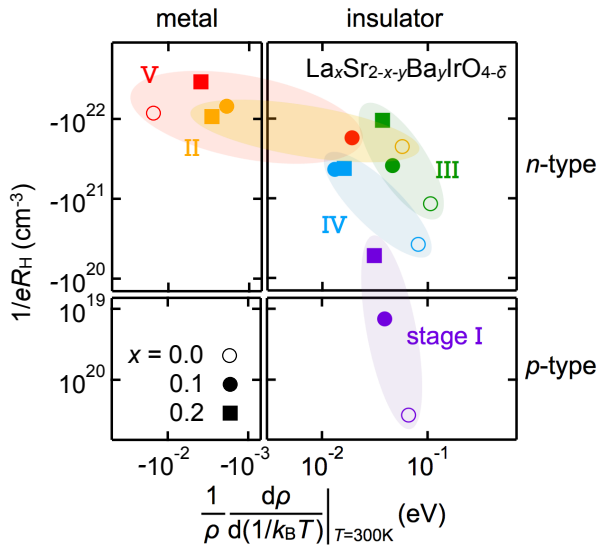}
\caption{(Color online)
Correlation between in-plane resistivity and Hall coefficient of the {\LSBIOd} films. The Ba doping content $y$ is set to 0.0 for the stages I, II, and III and 1.2 for IV and V. The horizontal axis indicates temperature differentiation of the resistivity at 300 K, which corresponds to the activation energy for insulating samples. The vertical axis indicates the inverse of the Hall coefficient $R_{\mathrm{H}}$, which represents the carrier density in one carrier model. Systematic changes from initial $p$-type insulator (lower right panel) to $n$-type metal (upper left panel) are confirmed through the stages.
}
\label{fig7}
\end{center}
\end{figure}

Finally we discuss relation between the in-plane resistivity and the carrier density of the films grown through the stages, as summarized in Fig. 7. Vertical axis indicates the carrier density calculated in one carrier model. The two upper and lower panels correspond to the electron and hole carriers, respectively. Horizontal axis shows temperature differentiation of the resistivity at 300 K, in the form equivalent to the activation energy as derived from Arrhenius plot for insulating samples. The right panels correspond to insulating films, while the left panels metallic ones. At the stage I, the system undergoes the change from $p$-type insulator to $n$-type insulator with increase of the La doping level. In stark contrast, even the non-doped film becomes $n$-type insulator by introducing the oxygen deficiency at the stage II, and then the $n$-type metallic state is realized by the La doping. At the stages III and IV, more straight Ir-O-Ir bond is tailored by the tensile substrate strain and the Ba substitution, although the system gets back to insulating again due to the lattice and/or valence mismatches to the {\GSO} substrate. At the stage V, more conducting $n$-type metal compared to the stage II is eventually achieved. These results demonstrate that it is important to design the Ir-O-Ir frame structure as well as to reduce the innate excess oxygen for the effective electron doping into the {\SIO} system.

\section{Conclusion}
In summary, we have fabricated {\LSBIOd} thin films by pulsed laser deposition and investigated the carrier doping effect with tailoring the in-plane Ir-O-Ir bond by tensile substrate strain and Ba substitution. The results demonstrate that it is important to make more straight Ir-O-Ir frame structure as well as to reduce innate excess oxygen for the effective electron doping. The relation between the charge transport and the carrier density has been elucidated through systematic changes from original $p$-type spin-orbit Mott insulator to highly doped $n$-type metal, which would be a good guide for realizing the superconductivity in the {\SIO} system.


\begin{acknowledgments}
This work was partly supported by Grant-in-Aids for Scientific Research (S) No. 24226002, Young Scientists (A) No. 15H05425, and Challenging Exploratory Research No. 26610098 from MEXT, Japan.
\end{acknowledgments}

\newpage
\end{document}